\documentclass[prb,reprint,twocolumn,showpacs,superscriptaddress,floatfix,aps,10pt]{revtex4-2}

\usepackage{amsmath, amsthm, amssymb}
\usepackage{dcolumn, bm, hyperref}
\usepackage{graphicx, subfigure, verbatim}
\usepackage{txfonts}    
\usepackage{newtxtext,newtxmath}
\usepackage{xcolor}
\usepackage{soul}
\usepackage{subfiles}
\usepackage{tabularx}

\DeclareMathOperator{\Tr}{Tr}

\usepackage{xcolor}

\begin{document}

\title{Intra-atomic magnetic octupoles and their coupling to cluster magnetic octupoles in Chiral antiferromagnets Mn$_3$Sn}

\author{Dohoon Park}
\affiliation{Department of Physics, Pohang University of Science and Technology, Pohang 37673, Korea}
\affiliation{Center for Quantum Dynamics of Angular Momentum, Pohang University of Science and Technology, Pohang 37673, Korea}
\author{Seungyun Han}
\email{hanson@postech.ac.kr}
\affiliation{Department of Physics, Pohang University of Science and Technology, Pohang 37673, Korea}
\affiliation{Center for Quantum Dynamics of Angular Momentum, Pohang University of Science and Technology, Pohang 37673, Korea}
\author{Hyun-Woo Lee}
\email{hwl@postech.ac.kr}
\affiliation{Department of Physics, Pohang University of Science and Technology, Pohang 37673, Korea}
\affiliation{Center for Quantum Dynamics of Angular Momentum, Pohang University of Science and Technology, Pohang 37673, Korea}

\begin{abstract}

We demonstrate that Mn$_3$Sn hosts finite intra-atomic magnetic octupoles (AMOs) $\mathbf{o}$ in addition to the well-established cluster magnetic octupole (CMO) $\mathbf{O}$. In contrast to the cluster-scale CMO, the AMO is a site-localized magnetic multipole associated with the anisotropic intra-atomic spin density. Symmetry analysis shows that the CMO and AMO transform in the same representation, allowing a bilinear interaction of the form $-g\,\mathbf{O}\cdot\mathbf{o}$. Using first-principles calculations, we confirm the presence of finite AMO densities and show that the AMO transforms concomitantly with the CMO under rotations of the noncollinear magnetic structure, providing microscopic evidence for the coupling between them. We further show that the magnetic band splitting can be represented by projected AMO operators, with manifold-dependent effective octupolar exchange coefficients in realistic Mn$_3$Sn. The presence of AMOs has three important implications. First, we show that the nonrelativistic spin splitting of Mn$_3$Sn can be described in terms of projected AMO operators, with manifold-dependent effective octupolar exchange coefficients, establishing the AMO as a microscopic operator underlying the spin splitting. Second, the AMO reveals a close connection between Mn$_3$Sn and $d$-wave altermagnets from the magnetic-octupole perspective. Third, the $\mathbf{O}\cdot\mathbf{o}$ coupling suggests a new route to manipulate the CMO using AMO currents, opening a direction for controlling the multipolar order in Mn$_3$Sn.

\end{abstract}

\maketitle

\section{Introduction}

The non-collinear antiferromagnet Mn$_3$Sn (Fig.~\ref{fig:1}) has attracted considerable attention due to its rich array of unconventional magnetic properties and transport phenomena \cite{nakatsuji2015}. Many of these have been attributed to cluster magnetic octupoles (CMOs), which emerge from its non-collinear spin structure and act as an effective order parameter~\cite{suzuki2017}. CMOs have been identified as the microscopic origin of the anomalous Hall effect (AHE) in Mn$_3$Sn and have been linked to various phenomena, including the anomalous Nernst effect (ANE) \cite{ikhlas2017, li2017} and magneto-optical Kerr effect (MOKE) \cite{higo2018}.


An intriguing aspect is that Mn$_3$Sn shares several key properties with $d$-wave altermagnets. These include nonrelativistic spin splitting \cite{hayami2019, yuan2020, yuan2021prm, smejkal2022symmetry, smejkal2022landscape, mazin2021}, large anomalous Hall conductivity even with vanishingly small net magnetization, tunneling magnetoresistance~\cite{dong2022}, and XMCD signals~\cite{yamasaki2020, kimata2021} . In $d$-wave altermagnets, these properties are understood to originate from a site-localized atomic magnetic octupole (AMO). The presence of the AMO is symmetry-guaranteed through its coupling to the N\'eel vector, $\mathbf{N} \cdot \mathbf{o}$, which has been extensively studied~\cite{bhowal2024, mcclarty2024}. This naturally raises the question: Do similar AMOs exist in Mn$_3$Sn? If so, how are they related to the CMOs?

In this work, we demonstrate that Mn$_3$Sn hosts finite AMOs in its noncollinear magnetic state (Fig.~\ref{fig:2}). Our symmetry analysis shows that the CMO and AMO vector transform in the same representation, allowing a bilinear interaction of the form $-g\,\mathbf{O}\cdot\mathbf{o}$, where $\mathbf{O}$ and $\mathbf{o}$ denote the CMO and AMO, respectively. First-principles calculations identify finite AMOs and show that their transformation under global rotations of the magnetic structure follows the symmetry relation predicted for the CMO. The rotation analysis therefore provides a first-principles confirmation of the symmetry correspondence between the CMO and the relevant AMO degrees of freedom.

The AMO vector $\mathbf{o}=(o_x,o_y)$ is a single two-component order parameter, with each component constructed from symmetry-equivalent orbital--spin basis functions. This structure provides a useful symmetry-based correspondence with the AMO description of $d$-wave altermagnets. At the same time, the noncollinear magnetic order and vector chirality of Mn$_3$Sn distinguish its symmetry setting from that of conventional collinear altermagnets.

The symmetry-allowed CMO--AMO relation also suggests a possible route for manipulating Mn$_3$Sn with AMO currents. Recent studies have proposed AMO-current-induced control of the N\'eel vector in $d$-wave altermagnets through an $\mathbf{N}\cdot\mathbf{o}$ coupling~\cite{han2025octupoleHall,baek2025magoct, han2026deterministic}. By analogy, the corresponding CMO--AMO relation in Mn$_3$Sn provides a natural starting point for exploring AMO-current-induced manipulation. A quantitative investigation of the resulting nonequilibrium torque and its efficiency is an interesting direction for future work.

Overall, our results connect the cluster-level CMO to a site-localized AMO description of the same noncollinear magnetic state. The symmetry analysis identifies the bilinear CMO--AMO invariant, the first-principles calculations establish the corresponding AMO structure, and the band analysis characterizes the associated magnetic splitting through manifold-dependent effective octupolar exchange coefficients.

This paper is organized as follows. In Sec.~II, we introduce and clarify the definitions of the CMO and AMO. In Sec.~III, we identify the symmetry-allowed bilinear coupling between them. In Sec.~IV, we examine the AMO structure and its transformation using density functional theory calculations. In Sec.~V, we characterize the AMO-associated magnetic band splitting using a minimal model and a manifold-resolved analysis of realistic Mn$_3$Sn. Finally, we discuss the broader implications of our results in Sec.~VI.


\section{Cluster magnetic octupole versus intra-atomic magnetic octupole}

\begin{figure}[t] \centering
\includegraphics[width=0.5\textwidth]{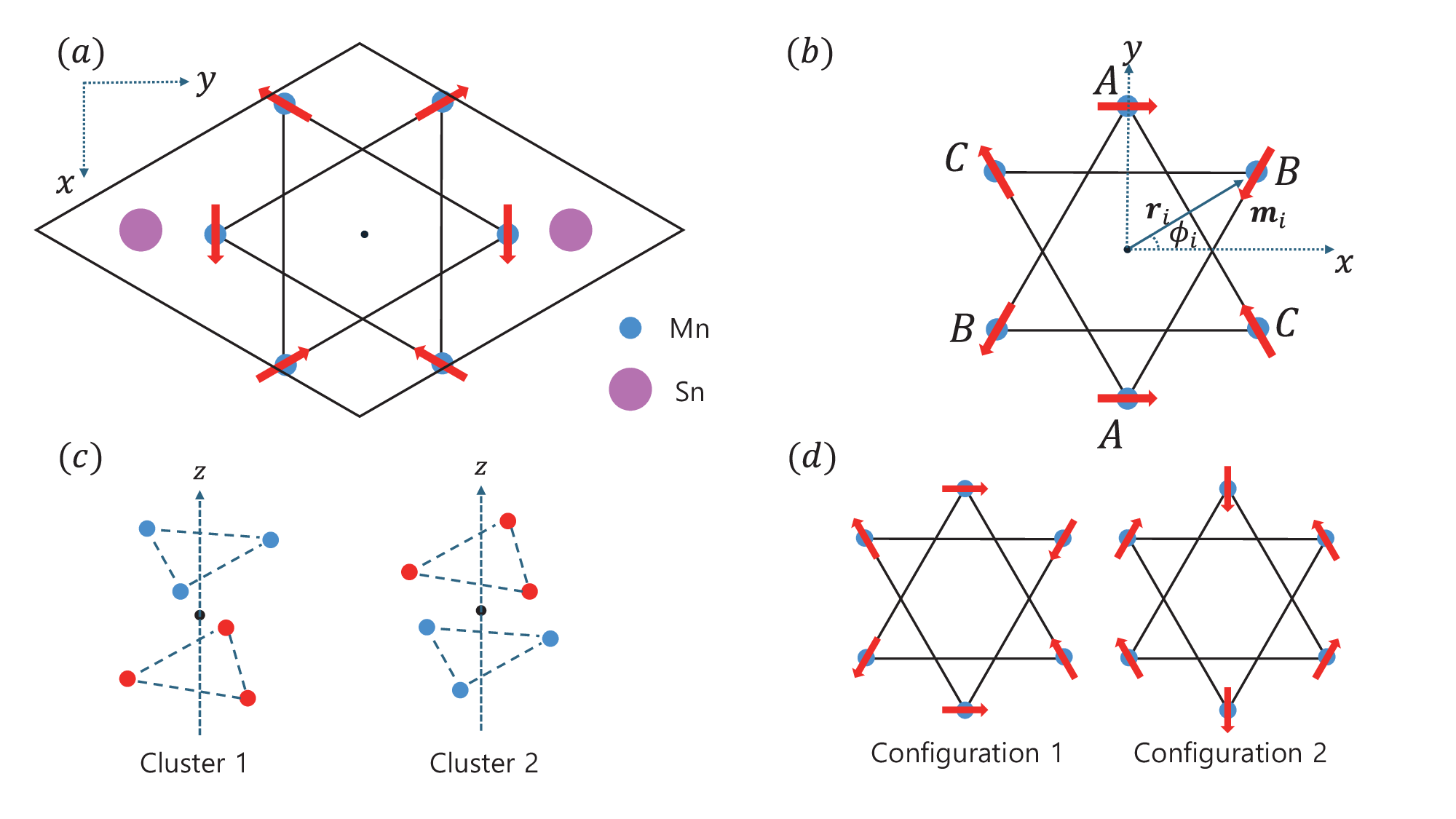}
\caption{(a) Unit cell of Mn$_3$Sn. (b) Visualization of the cluster multipole moment (CMO) in Mn$_3$Sn. (c) The Mn clusters can be categorized into two distinct types. (d) Two spin configurations are considered, each giving rise to negative vector chirality.}\label{fig:1} 
\end{figure}

\begin{figure}[t]
\includegraphics[width=0.5\textwidth]{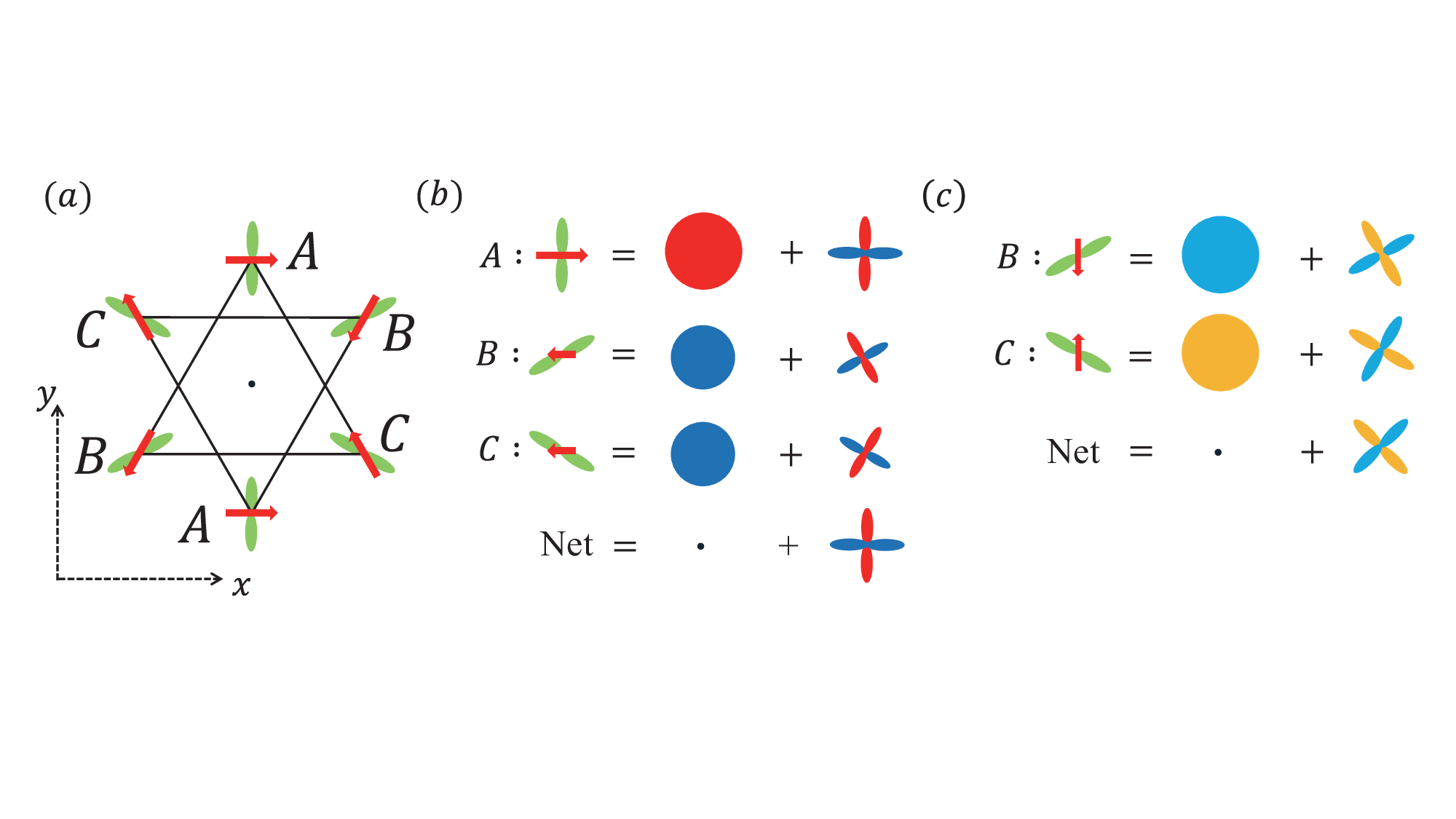}
\caption{(a) Intra-atomic magnetic multipoles at each Mn site in Mn$_3$Sn. The red arrows represent the magnetic moments of Mn atoms, and the green dumbbell shapes indicate the charge quadrupole density. (b) Multipole expansion for $S_x$ at the A, B, and C sites. The blue and red colors correspond to negative and positive $S_x$ densities, respectively. The resulting net AMO exhibits the $(L_x^2-L_y^2)S_x$ profile. (c) Multipole expansion for $S_x$ at the B, and C sites. The light blue and orange regions represent the negative and positive $S_y$ density. The resulting net AMO exhibits the $\{L_x,L_y\}S_y$ profile.}\label{fig:2} 
\end{figure}

Before proceeding to the main discussion, it is essential to clarify the definitions of the CMO and AMO, and to delineate the distinction between them. Both CMO and AMO deal with MO, which is defined as a third-rank tensor proportional to $r_n r_m S_l$.
The essential difference between the CMO and AMO lies in their spatial character: the former is \textit{intersite}, while the latter is \textit{intrasite}. For the CMO, $r_n$ represents the position vector measured from the cluster center to each sublattice site, Fig.~\ref{fig:1}(b).
Thus, the CMO corresponds to the MO formed by the spin arrangement of the sublattices (A, B, and C) around the cluster center.
In contrast, the AMO is defined within a single atomic site, where $r_n$ is measured from the center of the atomic site.
In this case, the MO arises from the local spin-density distribution, and the anisotropy of the spin-density profile plays a crucial role in determining the AMO, Fig.~\ref{fig:2}.

Now we review the CMO density in the Mn$_3$Sn \cite{suzuki2017}. The rank-$p$ cluster multipole (CMP) of the $\mu$-th cluster is defined as follows:
\begin{align} \label{clusterCMO}
    \mathbf{M}_{pq}^{(\mu)} = \sqrt{\frac{4\pi}{2p+1}}\sum_{i=1}^{N_{\mathrm{atom}}^{(\mu)}} \mathbf{m}_i \cdot \nabla_i (|\mathbf{r}_i|^pY_{pq}(\theta_i, \phi_i)^*),    
\end{align}
where $N_{\mathrm{atom}}^{(\mu)}$ is the number of atoms of the $\mu$-the cluster, $\mathbf{m}_i$ is the local magnetic moment of the $i$-th atom, $\nabla_i$ is the gradient with respect to the position vector $\mathbf{r}_i$ measured from the cluster center, $Y_{pq}$ is the spherical harmonics, and $\theta_i$ and $\phi_i$ are the polar and azimuthal angles of the $i$-th atom, respectively. The total CMP moment is defined as the sum over all clusters contained in a magnetic unit cell:
\begin{align} \label{CMO}
    \mathbf{M}_{pq} = \frac{N_{\mathrm{atom}}^\mathrm{u}}{N_{\mathrm{atom}}^\mathrm{c}} \frac{1}{V} \sum_{\mu = 1}^{N_\mathrm{cluster}} \mathbf{M}_{pq}^{(\mu)}
\end{align}
where $N_{\mathrm{atom}}^{\mathrm{u}}$ is the number of atoms in the magnetic unit cell, $N_{\mathrm{atom}}^{\mathrm{c}}$ is the total number of atoms in all magnetic clusters, $V$ is a volume of the magnetic unit cell, and $N_{\mathrm{cluster}}$ denotes the number of magnetic clusters in the unit cell. As an example, Mn$_3$Sn, which belongs to the $D_{6h}$ point group \cite{tomiyoshi1982, brown1990}, exhibits seven types of rank-$3$ CMPs ($p=3, q=\pm3, \pm2, \pm1, 0$) corresponding to this symmetry. 
Mn$_3$Sn comprises four clusters in total—two Mn clusters and two Sn clusters. Neglecting the magnetic moments on the Sn sites, the CMO of Mn$_3$Sn is therefore constructed solely from the two Mn clusters. In the present analysis, we adopt the spin configuration with negative vector chirality and, on this basis, proceed to evaluate the CMO. After defining the origin of each cluster and specifying the Mn-site positions, the CMO can be straightforwardly evaluated from Eqs.~\eqref{clusterCMO} and~\eqref{CMO}. Since the magnetic unit cell of Mn$_3$Sn contains six Mn atoms forming two Mn clusters, we have $N_{\mathrm{atom}}^{\mathrm{u}}=6$ and $N_{\mathrm{atom}}^{\mathrm{c}}=12$. The Mn clusters can be categorized into two types: Mn-cluster 1, in which the inverted triangle is located below, and Mn-cluster 2, in which it is located above [Fig.~\ref{fig:1}(c)]. In both cases, the origin of each cluster is defined as the midpoint between the centers of mass of the upright and inverted triangles. The types of irreducible representations are determined by the point-group symmetry of the system, and the CMPs are classified according to their corresponding irreducible representations. The CMO was evaluated for two spin configurations of the Mn atom in Fig.~\ref{fig:1}(d): configuration 1, where the spin at site A is aligned along the $x$-axis, and configuration 2, where the spin at site A is aligned along the $-y$-axis [Fig.~\ref{fig:1}(c)]. In these configurations, only two components of CMO, $\mathbf{M}_{3,1}$ and $\mathbf{M}_{3,-1}$, are non-zero. We combine them to define $\mathbf{O}_x$ and $\mathbf{O}_y$ as follows,
\begin{align}
    \mathbf{O}_x &\equiv \frac{1}{\sqrt{2}}(-\mathbf{M}_{3,1} + \mathbf{M}_{3,-1}) , \\
    \mathbf{O}_y &\equiv -\frac{i}{\sqrt{2}}(\mathbf{M}_{3,1} + \mathbf{M}_{3,-1}), 
\end{align} 
where $\mathbf{O}_x$ becomes finite in configuration 1 and vanishes in configuration 2, while $\mathbf{O}_y$ becomes finite in configuration 2 and vanishes in configuration 1. The magnitudes of both quantities are proportional to the magnetic moment of Mn. 
\\
Now we turn to the AMO. Unlike the CMO, which is defined over an atomic cluster, the AMO is defined at each atomic site. In Mn$_3$Sn, the net spin moment vanishes upon summation over sites. However, due to the distinct crystal-field environments at each site, the \textit{local} spin-density distribution becomes anisotropic, Fig.~\ref{fig:2}(a). As a result, a finite AMO can emerge at each site despite the absence of a net magnetic moment.
AMO is formally defined as follows ~\cite{jackeli2009, iwazaki2023, han2025octupoleHall}:
\begin{align}\label{generalAMO}
    {o}_{ij}^{{S}_k} = \frac{1}{\hbar^2} \left \{ {L}_i , {L}_j\right \}{S}_k 
\end{align}
where $\left \{ , \right \}$ is an anticommutator , $\left \{ A, B\right \} = AB + BA$, and $L_i, S_j$ are orbital angular momentum operator and spin angular momentum operator, respectively. Here $L_i$ captures the orbital angular momentum near each atom (atomic orbital angular momentum). Thus in Eq.~\eqref{generalAMO}, the anticommutator term of orbital angular momentum operators corresponds to the quadrupole density of electrons capturing the anisotropy of the local electronic density. To provide an intuitive picture, Fig.~\ref{fig:2}(a) illustrates a representative spin-density profile of Mn$_3$Sn. For the spin polarization along the $x$ direction, although the net spin moment vanishes upon summation over sites, the quantity $(L_x^2 - L_y^2)S_x$ remains finite. 
This indicates the presence of a nonzero AMO density proportional to $(x^2 - y^2)S_x$, as shown in Fig.~\ref{fig:2}(b). Similarly, for the spin polarization along the $y$ direction, the component $\{L_x, L_y\}S_y$ is finite, corresponding to an AMO density proportional to $xy S_y$.

In summary, the essential distinction between the CMO and the AMO lies in their definitions: the CMO is defined for an atomic cluster, as expressed in Eq.~\eqref{CMO}, whereas the AMO is defined by the spin density profile near each atomic site. In the following sections, we discuss the coupling between the CMO and the AMO.
%


\section{Landau theory analysis}

In this section, we determine whether a bilinear coupling between the CMO and AMO is allowed by the spin--space symmetry of Mn$_3$Sn. We identify the AMO degrees of freedom that transform in the same representation as the CMO and can therefore enter a bilinear invariant. This symmetry analysis identifies the allowed form of the interaction and the AMO components that can participate in it. We follow the symmetry-based procedure outlined in Ref.~\cite{mcclarty2024}.

In the limit of vanishing spin–orbit coupling, where the Dzyaloshinskii–Moriya interaction and magnetic anisotropy energies are negligible, we analyze Mn$_3$Sn within the framework of the spin space group symmetry. 
Within the $D_{6h}$ point group symmetry of Mn$_3$Sn, these CMOs transform as the $E_{1g}$ irreducible representation, behaving analogously to a conventional magnetic dipole vector. 
Rather than using CMO definition in Eq.~\eqref{CMO}, we use the alternative but equivalent definition of the CMO vectors which can be directly constructed based on the three Mn sublattices spin orientations as follows \cite{suzuki2017, nomoto2020}:
\begin{align}
    \mathbf{O}(\mathbf{S}^A,\mathbf{S}^B,\mathbf{S}^C)= \frac{1}{\sqrt{3}}{M_{xz}}\left[\mathbf{S}^A+R_z\left(\frac{2\pi}{3}\right)\mathbf{S}^B+R_z\left(-\frac{2\pi}{3}\right)\mathbf{S}^C \right],
\end{align}
where $\mathbf{O}(\mathbf{S}^A,\mathbf{S}^B,\mathbf{S}^C)$ is a vector made out of three site-spin moments of Mn$_3$Sn and $M_{xz}$ is an operator that inverts the $y$-component of the CMO. This CMO is invariant under the combined spin–space operation of  $[U_{3z} ||C_{6z}]$, but transforms non-trivially under the individual operations $C_{6z}$ (spatial) and $U_{3z}$ (spin) when applied separately. Specifically, under the $C_{6z}$ spatial rotation, the Mn sublattices transform cyclically as A $\rightarrow$ C, C $\rightarrow$ B, and B $\rightarrow$ A and the spin rotation $U_{3z}$ rotates each spin by 2$\pi$/3 about the $z$-axis. Each operation is represented as follows: 
%
\begin{align}
    \mathbf{O}\left (C_{6z} \left \{ \mathbf{S}^A,\mathbf{S}^B,\mathbf{S}^C \right \} \right )&=\mathbf{O}(\mathbf{S}^B,\mathbf{S}^C,\mathbf{S}^A) 
    \\&= R_z\left(\frac{2\pi}{3} \right)\mathbf{O}\left ( \mathbf{S}^A,\mathbf{S}^B,\mathbf{S}^C \right ), \\
    \mathbf{O} \left ( U_{3z}\left \{\mathbf{S}^A,\mathbf{S}^B,\mathbf{S}^C \right \} \right )&= R_z\left(-\frac{2\pi}{3} \right)\mathbf{O}\left ( \mathbf{S}^A,\mathbf{S}^B,\mathbf{S}^C \right ).
\end{align}
This verifies that the combined operation $[U_{3z} ||C_{6z}]$ leaves the CMO vector $\mathbf{O}(\mathbf{S}^A,\mathbf{S}^B,\mathbf{S}^C)$ invariant. Thus, the CMO belongs to a non-trivial irreducible representation under both $C_{6z}$ and $U_{3z}$ individually, yet remains invariant under their combined operation. This transformation property is analogous to that of the N\'eel vector in collinear AMs, which also transforms as a non-trivial representation under space-rotational symmetries but invariant under combined action with $U_2$ on spin which is a key feature of altermagnetism. 

Next, we identify the AMO that can couple to the CMO in Mn$_3$Sn. For a symmetry-allowed coupling to exist, the AMO must transform identically to the CMO under both spatial and spin operations. To this end, we construct a set of AMO vectors defined as follows \cite{jackeli2009, iwazaki2023, han2025octupoleHall}:
\begin{align}\label{AMO}
    o_x &= \left(L_x^2-L_y^2\right)S_x+\left \{L_x,L_y\right\} S_y, \nonumber\\
    o_y &= -\left(L_x^2-L_y^2\right)S_y + \left \{L_x,L_y\right \} S_x.
\end{align}
\begin{figure*}[t]
\includegraphics[width=1.0\textwidth]{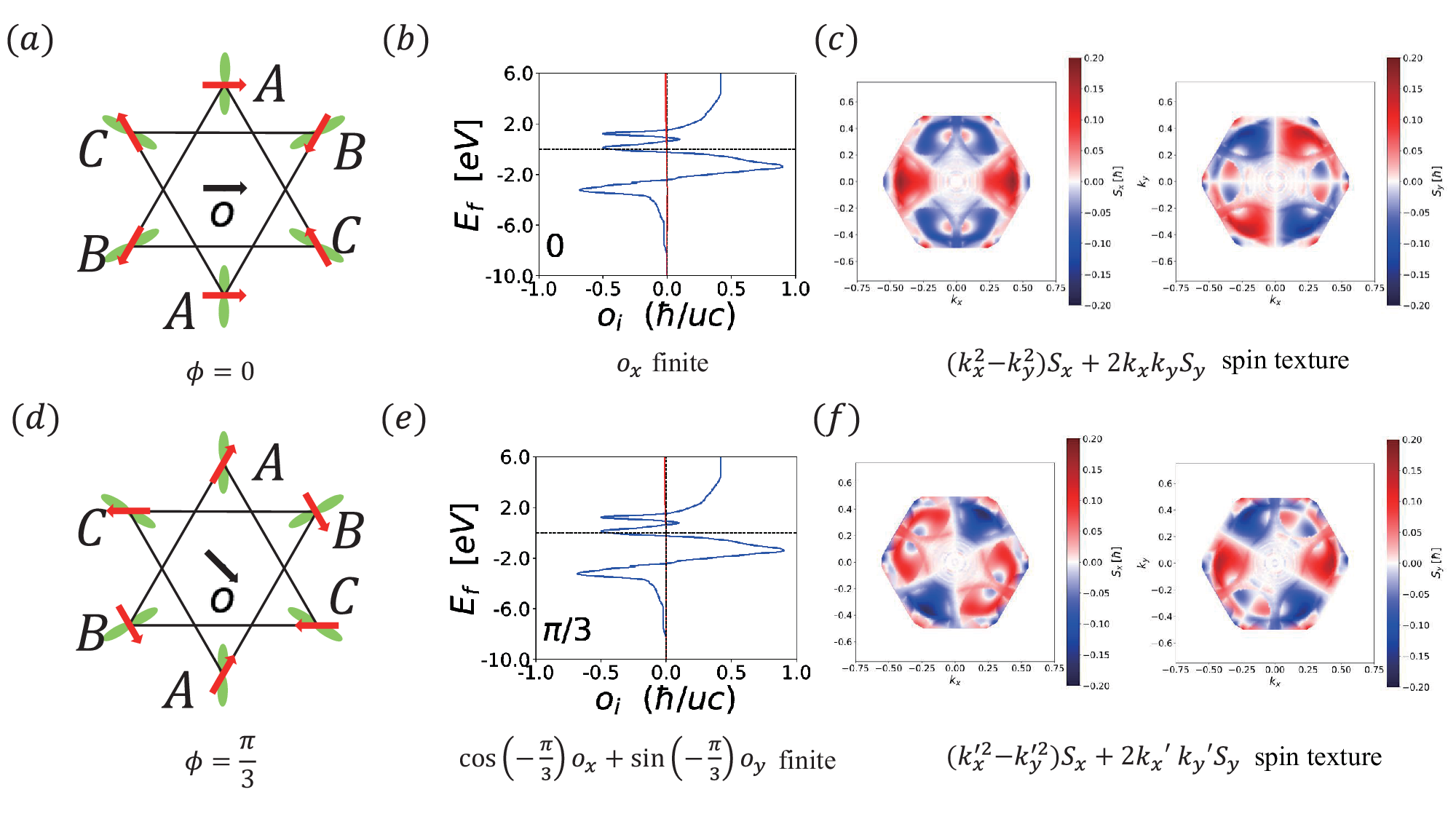}
\caption{(a) and (d) Spin configurations for $\phi=0$ and $\phi=\pi/3$, respectively.
The CMO in (a) is aligned parallel to the $x$-axis, while that in (d) is rotated clockwise by $\pi/3$ from the $x$-axis.
(b) and (e) Calculated density profiles for the two configurations.
Panel (b) for configuration (a) shows the densities of $(L_x^2-L_y^2)S_x+\{L_x,L_y\}S_y$ (blue) and $-(L_x^2-L_y^2)S_y+\{L_x, L_y\}S_x$ (red). Panel (e) for configuration (d) shows the densities of $\mathrm{cos}(-\frac{\pi}{3})(L_x^2-L_y^2)S_x+\mathrm{sin}(-\frac{\pi}{3})\{L_x,L_y\}S_y$ (blue) and $-\mathrm{sin}(-\frac{\pi}{3})(L_x^2-L_y^2)S_x+\mathrm{cos}(-\frac{\pi}{3})\{L_x,L_y\}S_y$ (red). (c) and (f) $S_x$ and $S_y$ spin textures in $k$-space corresponding to configurations (a) and (d), respectively.}\label{fig:3} 
\end{figure*}The components $o_x$ and $o_y$ are constructed from two symmetry-equivalent orbital--spin basis functions, $(L_x^2 - L_y^2)\mathbf{S}$ and $\{L_x,L_y\}\mathbf{S}$, and together form a two-dimensional AMO vector. Under $C_{6z}$ and $U_{3z}$, the AMO vector $\mathbf{o} = (o_x, o_y)$ transforms as follows:
\begin{align}
    C_{6z}\mathbf{o} &= R_z\left(\frac{2\pi}{3} \right)\mathbf{o}, \\
    U_{3z}\mathbf{o}&= R_z\left(-\frac{2\pi}{3} \right)\mathbf{o}.
\end{align}
This shows that this AMO vector belongs to the same irreducible representation as the CMO vector and thus satisfies the symmetry conditions required for bilinear coupling. That is, it transforms in the same way as the CMO vector under both spatial ($C_{6z}$) and spin ($U_{3z}$) operations and remains invariant under their combined action. Thus, there exists a symmetry-allowed coupling between the CMO and AMO vectors, which can be expressed as
\begin{equation}\label{coupling}
    H_{\text{Mn$_3$Sn}}\propto -g\,\mathbf{O}\cdot\mathbf{o},
\end{equation}

where \(g\) denotes the phenomenological coefficient of the symmetry-allowed bilinear invariant. The common transformation properties of $\mathbf{O}$ and $\mathbf{o}$ establish the bilinear CMO--AMO channel relevant to Mn$_3$Sn. Importantly, $\mathbf{o}=(o_x,o_y)$ forms a single two-component AMO order parameter, while the orbital--spin combinations entering Eq.~\eqref{AMO} act as symmetry-equivalent basis functions of this vector.

\section{Density functional calculations}

In this section, we use first-principles density functional theory (DFT) calculations to examine the AMO structure identified by the symmetry analysis and its connection to nonrelativistic spin splitting in Mn$_3$Sn. We first determine whether finite AMOs appear with the symmetry expected from the CMO configuration. Taking the configuration in which the local magnetic moment at the A site is aligned along the $+x$-axis as $\phi=0$, we systematically rotate the spin orientation at each site in a counterclockwise manner, thereby generating a series of CMO configurations. It is well known that such a rotation of 
$\phi$ induces a clockwise rotation of the CMO \cite{yoon2023handedness}. Building on this established behavior, we find that the AMO vector also rotates clockwise under the same variation of $\phi$, in accordance with the symmetry analysis discussed above. Specifically, we vary $\phi$ from $0$ to $\pi$ in steps of $\pi/6$ and obtain consistent results throughout this range. Among these configurations, $\phi=0, \pi/2,$ and $2\pi/3$ correspond to high-symmetry cases, for which the AMO structure and the associated $k$-space spin splitting are unambiguously dictated by symmetry. To highlight the behavior at intermediate angles, we therefore focus on the representative case of $\phi=\pi/3$ and directly compare it with the $\phi=0$ configuration.

For each configuration, we perform two steps of analysis: 1) we compute the AMO density using the AMO operators \cite{jackeli2009, iwazaki2023, han2025octupoleHall}, and 2) we confirm that the resulting real-space AMO configuration gives rise to a momentum-space spin splitting that is consistent with the AMO symmetry. For the first step, we perform the DFT calculation for hcp Mn$_3$Sn and obtain the electronic structures self-consistently using \textsc{fleur} \cite{fleurWeb}. The calculations were carried out using the generalized gradient approximation (GGA) in the form of the Perdew-Burke-Ernzerhof (PBE) exchange-correlation functional \cite{perdew1996}. We  employ a plane-wave cutoff of $k_\mathrm{{max}} = 4.5$ a.u.$^{-1}$ to expand the full-potential linearized augmented plane wave (LAPW) basis functions. The charge density converges using a Monkhorst-Pack \cite{monkhorst1976} $8 \times 8 \times 8$ $k$-point mesh over the full Brillouin zone. The muffin-tin radii are chosen to be $2.10$ a.u. for Mn and $2.30$ a.u. for Sn. The plane-wave cutoff parameters for the potential and the exchange-correlation potential are set to $g_\mathrm{{max}}=15.0$ a.u.$^{-1}$, and $g_\mathrm{{max,xc}}=12.5$ a.u.$^{-1}$, respectively. The lattice constants for hcp Mn$_3$Sn are $a = 5.66, c = 4.53$ \cite{zimmer1972}.

The DFT Bloch wave functions were projected onto maximally localized Wannier functions (MLWFs) \cite{mostofi2014} with the \textsc{wannier90} code \cite{pizzi2020wannier90}. The tight-binding (TB) Hamiltonian is constructed using atomic orbital-like MLWFs corresponding to Mn-$d$ and Sn-$s, p$ orbitals, capturing the electronic structure faithfully within an energy window of $2.5$ eV around the Fermi level.


The AMO density was calculated using the following AMO operators \cite{jackeli2009, iwazaki2023, han2025octupoleHall}:
\begin{align}
    \hat{o}_x &= \frac{1}{\hbar^2} \left[ \left(L_x^2-L_y^2\right)S_x+\left \{L_x,L_y\right\} S_y \right], \\
    \hat{o}_y &= \frac{1}{\hbar^2} \left[ -\left(L_x^2-L_y^2\right)S_y + \left \{L_x,L_y\right \} S_x \right],
\end{align}
where $L_i$ and $S_i$ are the orbital angular momentum and spin operators and $\left\{ L_x, L_y \right \}$ is an anti-commutator of $L_x$ and $L_y$. Here, $L_x^2-L_y^2$ and $\left \{L_x, L_y \right \}$ represent the electronic quadrupole density as described using the spherical tensors and the generalized Stevens operator approach $\left \{L_i, L_j  \right \}/\hbar^2=Cr_ir_j/a_0^2$, where $C$ is constant and $a_0$ is the Bohr radius \cite{hayami2024unified, shiina1997, kuramoto2009, kusunose2008, santini2009}. Then the AMO density is obtained as follows:
\begin{align}\label{density}
    {o}_{i} &= \frac{1}{NV_\mathrm{{unit}}} \sum_{n \mathbf{k}} f_{n \mathbf{k}} \langle u_{n\mathbf{k}}| \hat{o_i}| u_{n\mathbf{k}} \rangle, \nonumber \\
    {o}_{i,\mathrm{unit}} &= \frac{1}{N} \sum_{n \mathbf{k}} f_{n \mathbf{k}} \langle u_{n\mathbf{k}}| \hat{o_i}| u_{n\mathbf{k}} \rangle,
\end{align}
where $i$ denotes either $x$ or $y$, and $N$ is the number of points in the $k-$grid, which is $80 \times80\times80$ in this case. $V_\mathrm{{unit}}$ represents the unit cell volume, and $f_{n\mathbf{k}}$ refers to the Fermi–Dirac distribution function. All calculations were carried out at room temperature. Equation~\eqref{density} gives the AMO density per unit cell.

We then examined how the AMO density changes when the CMO direction is rotated. For $\phi=0$, corresponding to $\mathbf{O}$ aligned along the $x$ direction, only the $o_x$ density is finite [Fig.~\ref{fig:3}(b)], whereas $o_y$ vanishes. For $\phi=\pi/3$, the finite AMO follows the corresponding rotated direction [Fig.~\ref{fig:3}(e)]. This transformation agrees with the symmetry analysis and confirms that the AMO degrees of freedom identified in Sec.~III correctly describe the first-principles results. Together with the finite AMO obtained for each magnetic configuration, the systematic evolution of its orientation provides a consistent first-principles confirmation of the symmetry-allowed bilinear CMO--AMO relation.

%
%
Second, we examine the correspondence between the finite AMO and the nonrelativistic $k$-space spin splitting.
This can be directly verified through the mapping previously introduced in Ref.~\cite{bhowal2024}. In that study, the relationship between the AMO and $k$-space spin splitting was established via a tensor mapping of the form $r_n r_m S_l \rightarrow k_n k_m S_l$ near the $\Gamma$ point, demonstrating that the real-space AMO gives rise to an equivalent momentum-space spin texture. Here we show that this mapping also holds in Mn$_3$Sn. As shown in Fig.~\ref{fig:3}, when $\phi=0$ and $o_x$ is finite, the spin texture in momentum space exhibits the form $(k_x^2 - k_y^2) S_x + 2 k_x k_y S_y$ [Fig.~\ref{fig:3}(c)] near the $\Gamma$ point. For $\phi=\pi/3$, where $\mathrm{cos}(-\frac{\pi}{3})o_x + \mathrm{sin}(-\frac{\pi}{3})o_y$ becomes finite, the spin splitting takes the form $-(k_x'^2 - k_y'^2) S_y + 2 k_x' k_y' S_x$ near the $\Gamma$ point, where $(k_x', k_y')$ denotes the coordinate system obtained by rotating $(k_x, k_y)$ by $\frac{\pi}{3}$, consistent with the expected mapping of the real-space AMO structure to momentum space. These results establish the expected correspondence between the real-space AMO structure and the nonrelativistic momentum-space spin texture.

\begin{figure*}[t]
    \centering
    \includegraphics[width=\textwidth]{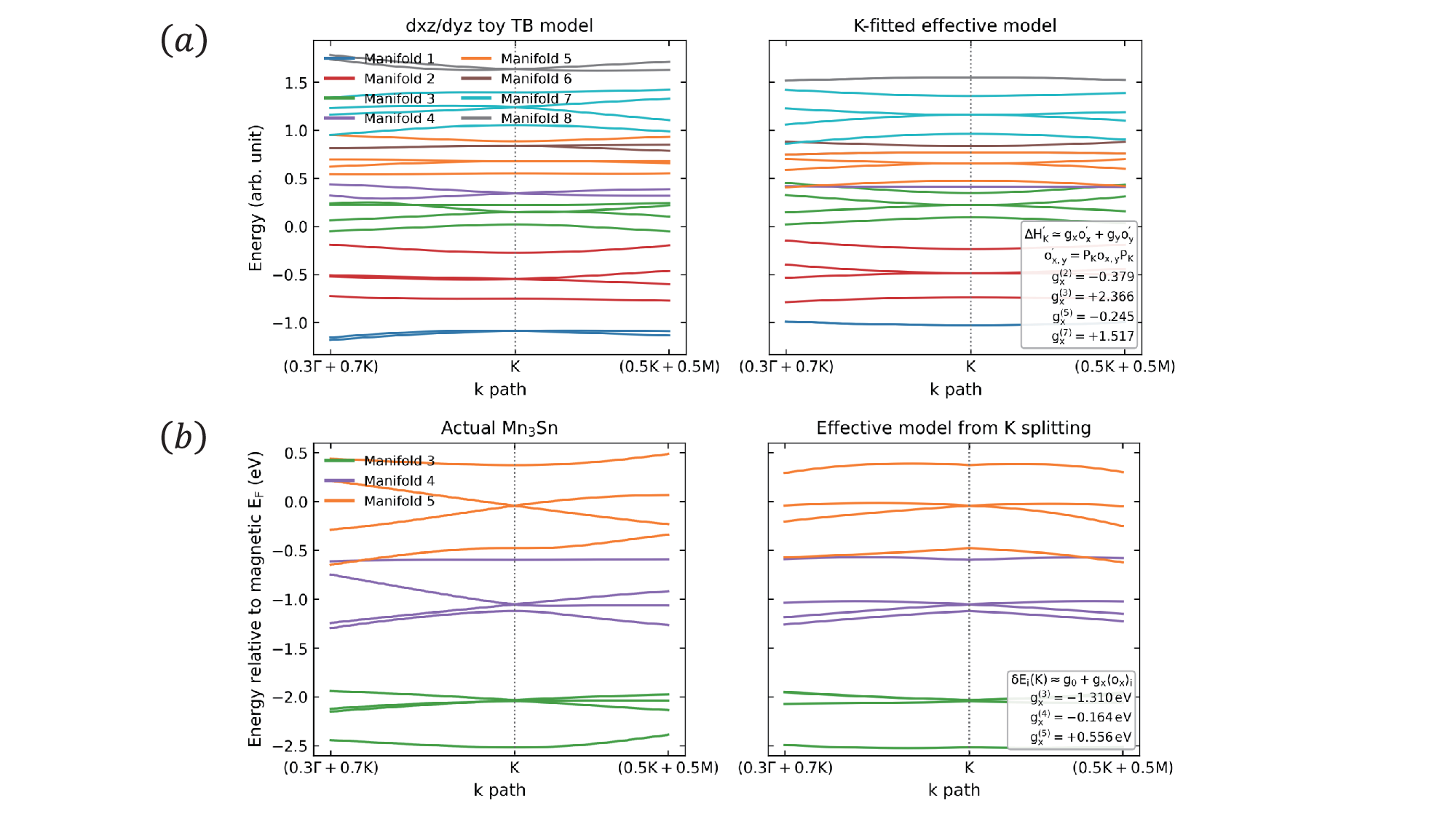}
    \caption{
    (Color online) Full and effective band structures along
    $(0.3\Gamma+0.7K)-K-(0.5K+0.5M)$.
    Colors label the parent manifolds at $K$, and the dotted vertical line marks the $K$ point.
    In (a), the left and right panels show the exact and effective bands of the minimal
    $d_{xz}/d_{yz}$ toy tight-binding model, respectively.
    In (b), the left and right panels show the actual and effective bands for Mn$_3$Sn, respectively.
    The good agreement near $K$ shows that the magnetic splitting is captured by the
    manifold-resolved octupolar response.
    }
    \label{fig:fig4}
\end{figure*}

\section{Manifold-resolved effective octupolar exchange coefficients}

In this section, we characterize the AMO-associated magnetic band splitting in the CMO-ordered state. The coefficient $g$ in Eq.~\eqref{coupling} denotes the phenomenological coefficient of the symmetry-allowed bilinear invariant,
\begin{align}
    H_{\mathrm{int}} = - g\,\mathbf{O}\cdot\mathbf{o}.
\end{align}

The quantities extracted below are manifold-resolved coefficients of an effective single-particle description. Throughout this analysis, the CMO amplitude is fixed and normalized in our convention, so the fitted coefficients characterize the octupolar band response in this fixed CMO background. Since the ground-state CMO considered here is oriented along $O_x$, the relevant projected AMO channel is $o_x$. Our goal is to determine whether the magnetic band splitting of Mn$_3$Sn near the $K$ point can be represented by projected AMO operators and to extract the corresponding manifold-dependent effective octupolar exchange coefficients. We focus on the coupling near the $K$ point for two reasons. First, in Mn$_3$Sn, the electronic states near the $K$ point play an important role in low-energy transport phenomena, and the magnetic splitting in this region directly affects the electronic structure relevant to transport. Second, the $K$ point provides a natural reference point for identifying the magnetic splitting: in the nonmagnetic reference system, several bands form well-defined nearly degenerate groups near the $K$ point, while the noncollinear magnetic order splits these groups into magnetic child bands.

This allows us to characterize the octupolar band response in a controlled, manifold-resolved manner. To be specific, we consider the following effective single-particle form, 
\begin{align}
    H_{\mathrm{eff}}^{\mathrm{AMO}}
    =
    -\sum_m g_x^{(m)} O_x\, o_x^{(m)},
\end{align}
where $m$ labels a parent manifold at the $K$ point and $o_x^{(m)}$ denotes the AMO operator projected onto that manifold. Here, a parent manifold refers to a set of nonmagnetic bands at the $K$ point that are close in energy and share the same dominant orbital character. When the noncollinear magnetic order is introduced, each parent manifold splits into a set of magnetic child states. We then ask whether the splitting pattern of these child states can be explained by the expectation value of the AMO operator within the corresponding parent manifold. 
This manifold-resolved treatment is essential because the AMO is not a purely spin quantity. 
It contains the orbital quadrupole operators, such as $L_x^2-L_y^2$ and $\{L_x,L_y\}$, and therefore depends sensitively on the anisotropy of the electronic density. 
Since different bands have different orbital composition and different local density anisotropy, the effective octupolar exchange coefficient is naturally manifold dependent.

Before applying this procedure to realistic Mn$_3$Sn, we first use a minimal tight-binding model to clarify the role of the AMO coupling. 
The purpose of the model is not to reproduce the full band structure of Mn$_3$Sn quantitatively, but to demonstrate at the Hamiltonian level that the magnetic splitting near $K$ can indeed be represented by the projected AMO operators. 
This is important because the realistic band comparison alone shows that an effective model can reproduce the bands, but it does not by itself establish why the effective term should be identified with the AMO coupling. 
The minimal model provides this microscopic link by explicitly constructing both the magnetic correction and the AMO operators in the same Hilbert space.

\subsection{Minimal $d_{xz}/d_{yz}$ tight-binding model}

We consider a minimal tight-binding model defined on the six Mn sites of the Mn$_3$Sn unit cell, retaining only two real $d$ orbitals, $(d_{xz},d_{yz})$, on each Mn site. 
Including spin, the Hilbert-space dimension is
\begin{align}
    N_{\mathrm{Hilbert}} = 6 \times 2 \times 2 = 24.
\end{align}
The Hamiltonian is written as
\begin{align}
    H_{\mathrm{TB}}(\mathbf{k})
    =
    H_{\mathrm{hop}}(\mathbf{k})
    +H_{\mathrm{cf}}
    +H_{\mathrm{ex}} .
\end{align}
The onsite crystal-field term $H_{cf}$ in the $(d_{xz},d_{yz})$ basis is
\begin{align}
    H_{\mathrm{cf}}
    =
    \sum_i
    \begin{pmatrix}
        \Delta_{xz} & 0 \\
        0 & \Delta_{yz}
    \end{pmatrix}_i,
\end{align}
with
\begin{align}
    \Delta_{xz}=0,\qquad \Delta_{yz}=0.8.
\end{align}
The hopping term $H_{\mathrm{hop}}(\mathbf{k})$is constructed from the Slater--Koster-type $d$--$d$ hopping amplitudes projected onto the $(d_{xz},d_{yz})$ sector. 
For a bond direction $\hat{\mathbf d}=(l,m,n)$, the hopping matrix is parameterized by $V_{dd\sigma}$, $V_{dd\pi}$, and $V_{dd\delta}$. 
In the present calculation, we use the nearest-neighbor values
\begin{align}
    V_{dd\sigma}^{(1)} = 0.4,\qquad
    V_{dd\pi}^{(1)}    = -0.2,\qquad
    V_{dd\delta}^{(1)} = 0.1.
\end{align}
The noncollinear magnetic order is modeled by a local exchange field,
\begin{align}
    H_{\mathrm{ex}}
    =
    -J_{\mathrm{ex}}\sum_i \mathbf{S}_i \cdot \boldsymbol{\sigma},
\end{align}
where $\boldsymbol{\sigma}$ is the vector of Pauli matrices and $\mathbf{S}_i$ forms the coplanar $120^\circ$ magnetic texture with $J_{\mathrm{ex}} = 0.1$.

Within this Hilbert space, the two symmetry-allowed AMO operators are
\begin{align}
    \hat{o}_x
    &=
    (L_x^2-L_y^2)S_x+\{L_x,L_y\}S_y, \\
    \hat{o}_y
    &=
    -(L_x^2-L_y^2)S_y+\{L_x,L_y\}S_x .
\end{align}
We analyze the magnetic splitting at the $K$ point by comparing the nonmagnetic and magnetic Hamiltonians,
\begin{align}
    H_{\mathrm{off}}(K) &= H_{\mathrm{TB}}(K)\big|_{J_{\mathrm{ex}}=0},\\
    H_{\mathrm{on}}(K)  &= H_{\mathrm{TB}}(K)\big|_{J_{\mathrm{ex}}\neq 0},
\end{align}
and define the magnetic correction
\begin{align}
    \Delta H_K = H_{\mathrm{on}}(K)-H_{\mathrm{off}}(K).
\end{align}

Let $P_m$ be the projector onto the $m$-th parent manifold of the nonmagnetic Hamiltonian at $K$. 
The projected magnetic correction and projected AMO operators are then defined as
\begin{align}
    \Delta H_K^{(m)}
    &=
    P_m \Delta H_K P_m,\\
    o_x^{(m)}
    &=
    P_m \hat{o}_x P_m,\\
    o_y^{(m)}
    &=
    P_m \hat{o}_y P_m.
\end{align}
Here, $\Delta H_K^{(m)}$ is the magnetic perturbation restricted to the $m$-th parent manifold. 
Its trace part shifts all states in the manifold equally and therefore does not contribute to the splitting pattern. 
We thus separate the traceless part as
\begin{align}
    \Delta \tilde{H}_K^{(m)}
    =
    \Delta H_K^{(m)}
    -
    \frac{\Tr[\Delta H_K^{(m)}]}{d_m}\,\mathbb{I}_m,
\end{align}
where $d_m=\Tr P_m$ is the dimension of the parent manifold and $\mathbb{I}_m$ is the identity matrix within that manifold. 
Similarly, we define the traceless projected AMO operators
\begin{align}
    \tilde{o}_\alpha^{(m)}
    =
    o_\alpha^{(m)}
    -
    \frac{\Tr[o_\alpha^{(m)}]}{d_m}\,\mathbb{I}_m,
    \qquad
    \alpha=x,y.
\end{align}
The effective coefficients are then obtained by fitting
\begin{align}
    \Delta \tilde{H}_K^{(m)}
    \simeq
    g_x^{(m)}\tilde{o}_x^{(m)}
    +
    g_y^{(m)}\tilde{o}_y^{(m)} .
    \label{eq:toy_fit}
\end{align}
This fit is performed using the Hermitian inner product
\begin{align}
    \langle A,B\rangle_{\mathrm{H}}
    =
    \Tr(A^\dagger B).
\end{align}
Equation~\eqref{eq:toy_fit} shows explicitly that the splitting of each parent manifold can be represented by the projected AMO operators. 
Thus, the minimal model demonstrates at the Hamiltonian level that the magnetic splitting can be represented by projected AMO operators, providing the basis for the effective description used below for realistic Mn$_3$Sn.

\subsection{Manifold-resolved extraction in Mn$_3$Sn}

We now apply the same idea to realistic Mn$_3$Sn. 
Starting from a nonmagnetic Wannier Hamiltonian, we identify the relevant bands at the $K$ point and group them into parent manifolds according to their energy proximity and dominant orbital character. 
For the $m$-th parent manifold, we define its reference energy as the average nonmagnetic energy,
\begin{align}
    E_0^{(m)}
    =
    \frac{1}{d_m}
    \sum_{a\in m} E_{a}^{\mathrm{off}}(K),
\end{align}
where $E_{a}^{\mathrm{off}}(K)$ are the nonmagnetic eigenvalues belonging to the parent manifold and $d_m$ is the number of nonmagnetic bands in that manifold.

When the noncollinear magnetic order is introduced, this parent manifold splits into magnetic child states. 
For the $i$-th child state originating from the $m$-th parent manifold, we define the magnetic energy shift
\begin{align}
    \delta E_i^{(m)}
    =
    E_i^{\mathrm{on}}(K)-E_0^{(m)} ,
\end{align}
where $E_i^{\mathrm{on}}(K)$ is the magnetic eigenvalue at $K$. 
We then relate this energy shift to the AMO expectation values of the corresponding magnetic eigenstate,
\begin{align}
    \delta E_i^{(m)}
    \simeq
    g_0^{(m)}
    +
    g_x^{(m)}
    \langle o_x\rangle_i^{(m)}
    +
    g_y^{(m)}
    \langle o_y\rangle_i^{(m)} .
    \label{eq:real_fit_general}
\end{align}
Here, $g_0^{(m)}$ describes the average energy shift of the manifold, whereas $g_x^{(m)}$ and $g_y^{(m)}$ describe the splitting associated with the AMO character of the magnetic child states.

In the configuration considered here, the CMO is aligned along the $O_x$ direction. 
Consistently, the calculated splitting is dominated by the $o_x$ channel, while the $o_y$ contribution is negligible. 
We therefore use the reduced fitting form
\begin{align}
    \delta E_i^{(m)}
    \simeq
    g_0^{(m)}
    +
    g_x^{(m)}
    \langle o_x\rangle_i^{(m)} .
    \label{eq:real_fit_ox}
\end{align}

The extracted $g_x^{(m)}$ defines a manifold-dependent effective octupolar exchange coefficient within the $m$-th parent manifold near the $K$ point.

We emphasize that $g_x^{(m)}$ is not expected to be the same for all manifolds. 
The AMO operator contains the orbital quadrupole part, and therefore its expectation value measures the anisotropy of the local electronic density together with the spin polarization. 
Since each band has a different orbital composition and a different anisotropic density profile, its response to the projected AMO operator is naturally band and manifold dependent. 

This naturally leads to a manifold-resolved description in terms of effective octupolar exchange coefficients.

To test the validity of this description beyond the fitting point itself, we construct a $K$-centered effective Hamiltonian by adding the extracted manifold-resolved splitting to the nonmagnetic reference Hamiltonian,
\begin{align}
    H_{\mathrm{eff}}(\mathbf{k})
    =
    H_{\mathrm{off}}(\mathbf{k})
    +
    \sum_m
    P_m
    \left[
        g_0^{(m)}
        +
        g_x^{(m)} o_x^{(m)}
    \right]
    P_m .
    \label{eq:effective_H}
\end{align}
This effective Hamiltonian is intended to describe the band structure in the vicinity of the $K$ point, where the parent manifolds are defined. 
It keeps the nonmagnetic dispersion away from $K$ while incorporating the AMO-induced magnetic splitting extracted at $K$.

\subsection{Numerical results}

The extracted effective octupolar exchange coefficients for realistic Mn$_3$Sn show strong manifold dependence near the Fermi level. 
Among the parent manifolds relevant to the low-energy electronic structure, the largest response is found for the third parent manifold,
\begin{align}
    g_x^{(3)} \approx -1.310~\mathrm{eV},
\end{align}
indicating strong octupolar sensitivity in this subspace. 
The fifth manifold also exhibits a sizable positive coefficient,
\begin{align}
    g_x^{(5)} \approx +0.556~\mathrm{eV},
\end{align}
whereas the fourth manifold shows a smaller but still visible negative value,
\begin{align}
    g_x^{(4)} \approx -0.164~\mathrm{eV}.
\end{align}

The sign of $g_x^{(m)}$ reflects the relative correspondence between the projected AMO operator and the magnetic band splitting within the corresponding manifold, while its magnitude quantifies the effective octupolar response of that manifold.

Figure~\ref{fig:fig4} summarizes the quality of this description near the Fermi level. 
For the minimal tight-binding model, the exact magnetic bands and the effective bands obtained from the projected AMO coupling agree well near the $K$ point, confirming that the AMO operators capture the magnetic splitting at the Hamiltonian level. 
For realistic Mn$_3$Sn, the effective model constructed from the extracted $g_x^{(m)}$ values reproduces the magnetic band splitting of the manifolds near the Fermi level. 
This agreement demonstrates that the low-energy band structure near $K$ is well described by a manifold-resolved effective octupolar single-particle term,
\begin{align}
    H_{\mathrm{eff}}^{\mathrm{AMO}}
    \sim
    -\sum_m g_x^{(m)} O_x\,o_x^{(m)} .
\end{align}
In particular, if we focus on the manifold at the Fermi energy (manifold $m=5$), which is most important for transport, we may retain only the coupling for $m=5$. 
\subsection{Physical implication}

The combined toy-model and realistic analyses lead to a unified physical picture. 

The nonrelativistic magnetic splitting near the $K$ point is not simply a uniform exchange splitting but follows the AMO character of the electronic states within each parent manifold. 
Because the AMO reflects the anisotropic intra-atomic electronic density, the corresponding effective octupolar response is naturally manifold dependent. 
Once this manifold dependence is included, the magnetic band structure near $K$ is accurately captured by the projected AMO description. 
This establishes a direct connection between the symmetry-allowed CMO--AMO channel and the manifold-resolved single-particle response of realistic Mn$_3$Sn. When one considers only the band manifold at the Fermi energy, the corresponding effective coefficient provides the relevant low-energy description.

\section{Discussion}

Our results reveal a direct symmetry connection between the cluster and intra-atomic magnetic octupoles in Mn$_3$Sn. The CMO and AMO transform in the same representation, allowing the bilinear invariant $-g\,\mathbf{O}\cdot\mathbf{o}$, while the first-principles calculations identify finite AMOs with the corresponding symmetry. The band analysis further shows that the associated magnetic splitting can be represented by projected AMO operators with manifold-dependent effective octupolar exchange coefficients. Together, these results connect the cluster-scale magnetic order to the local anisotropic spin density.

The AMO therefore provides a complementary local perspective on the noncollinear magnetic state. In particular, the calculated real-space AMO and momentum-space spin texture exhibit the expected symmetry correspondence, supporting an AMO-based description of the nonrelativistic spin splitting. This local multipolar perspective may also be useful for interpreting probes such as XMCD, which are sensitive to intra-atomic electronic and magnetic structure~\cite{yamasaki2020, kimata2021}, and for relating such local responses to the non-local CMO-based description of Mn$_3$Sn.


The AMO structure identified here also shows a magnetic multipole-based correspondence with $d$-wave altermagnets. This connection highlights why Mn$_3$Sn and $d$-wave altermagnets share similar properties despite distinct noncollinear magnetic order and vector chirality of Mn$_3$Sn.

Finally, we comment on a possible implication of the CMO--AMO relation.

Recent theoretical studies have proposed AMO-current-induced torques in $d$-wave altermagnets through a coupling between the N\'eel vector and the AMO~\cite{han2025octupoleHall, han2026deterministic}. By symmetry analogy, the CMO--AMO relation identified here suggests that AMO-current-induced manipulation may also be possible in Mn$_3$Sn. Quantifying the corresponding torque, transport response, and efficiency in Mn$_3$Sn will be an important direction for future work.

\section*{Acknowledgement}
We thank Insu Baek, Dongwook Go, and Hojun Lee for fruitful discussion. D.P., S.H., and H.-W.L. were financially supported by the National Research Foundation of Korea (NRF) grant funded by the Korean government (MSIT) (No. RS-2024-00356270 and RS-2024-00410027). Supercomputing resources, including technical support, were provided by the Supercomputing Center, Korea Institute of Science and Technology Information (Contract No. KSC-2025-CRE-0582). 

\section*{Data availability}
The data underlying the numerical results presented in the figures are available from the corresponding author upon reasonable request.

\bibliography{references_revised}

@article{yamasaki2020,
  title={Augmented magnetic octupole in {K}agom{\'e} 120-degree antiferromagnets detectable via {X}-ray magnetic circular dichroism},
  author={Yamasaki, Yuichi and Nakao, Hironori and Arima, Taka-hisa},
  journal={Journal of the Physical Society of Japan},
  volume={89},
  number={8},
  pages={083703},
  year={2020},
  publisher={The Physical Society of Japan}
}

@article{kimata2021, 
title={X-ray study of ferroic octupole order producing anomalous {H}all effect}, author={Kimata, Motoi and Sasabe, Norimasa and Kurita, Kensuke and Yamasaki, Yuichi and Tabata, Chihiro and Yokoyama, Yuichi and Kotani, Yoshinori and Ikhlas, Muhammad and Tomita, Takahiro and Amemiya, Kenta and others}, journal={Nature Communications}, volume={12}, number={1}, pages={5582}, year={2021}, publisher={Nature Publishing Group UK London} }

@article{nakatsuji2015,
  title={Large anomalous {H}all effect in a non-collinear antiferromagnet at room temperature},
  author={Nakatsuji, Satoru and Kiyohara, Naoki and Higo, Tomoya},
  journal={Nature},
  volume={527},
  number={7577},
  pages={212--215},
  year={2015},
  publisher={Nature Publishing Group}
}

@article{suzuki2017,
  title={Cluster multipole theory for anomalous {H}all effect in antiferromagnets},
  author={Suzuki, M. T. and Koretsune, T. and Ochi, M. and Arita, R.},
  journal={Physical Review B},
  volume={95},
  number={9},
  pages={094406},
  year={2017},
  publisher={American Physical Society}
}

@article{ikhlas2017,
  title={Large anomalous {N}ernst effect at room temperature in a chiral antiferromagnet},
  author={Ikhlas, M. and Tomita, T. and Koretsune, T. and Suzuki, M.-T. and Nishio-Hamane, D. and Arita, R. and Otani, Y. and Nakatsuji, S.},
  journal={Nature Physics},
  volume={13},
  pages={1085--1090},
  year={2017},
  publisher={Nature Publishing Group}
}

@article{li2017,
  title={Anomalous {N}ernst and {R}ighi-{L}educ effects in {Mn}$_3${Sn}: {B}erry curvature and entropy flow},
  author={Li, X. and Xu, L. and Ding, L. and Wang, J. and Shen, M. and others},
  journal={Physical Review Letters},
  volume={119},
  pages={056601},
  year={2017},
  publisher={American Physical Society}
}

@article{higo2018,
  title={Large magneto-optical {K}err effect and imaging of magnetic octupole domains in an antiferromagnetic metal},
  author={Higo, Tomoya and Man, Huiyuan and Gopman, Daniel B. and Wu, Liang and Koretsune, Takashi and Van 'T Erve, Olaf M. J. and others},
  journal={Nature Photonics},
  volume={12},
  number={2},
  pages={73--78},
  year={2018},
  publisher={Nature Publishing Group}
}

@article{dong2022,
  title={Tunneling {M}agnetoresistance in {N}oncollinear {A}ntiferromagnetic {T}unnel {J}unctions},
  author={Dong, Jianting and Li, Xinlu and Gurung, Gautam and Zhu, Meng and Zhang, Peina and Tsymbal, Evgeny Y. and Jia, Zhang},
  journal={Physical Review Letters},
  volume={128},
  number={19},
  pages={197201},
  year={2022},
  publisher={American Physical Society}
}

@article{hayami2019,
  title={Momentum-dependent spin splitting by collinear antiferromagnetic ordering},
  author={Hayami, Satoru and Yanagi, Yuki and Kusunose, Hiroaki},
  journal={Journal of the Physical Society of Japan},
  volume={88},
  pages={123702},
  year={2019},
  publisher={Physical Society of Japan}
}

@article{smejkal2022landscape,
  title={Emerging research landscape of altermagnetism},
  author={{\v{S}}mejkal, Libor and Sinova, Jairo and Jungwirth, Tomas},
  journal={Physical Review X},
  volume={12},
  pages={040501},
  year={2022},
  publisher={American Physical Society}
}

@article{yuan2020,
  title={Giant momentum-dependent spin splitting in centrosymmetric low-{Z} antiferromagnets},
  author={Yuan, L.-D. and Wang, Z. and Luo, J.-W. and Rashba, E. I. and Zunger, Alex},
  journal={Physical Review B},
  volume={102},
  pages={014422},
  year={2020},
  publisher={American Physical Society}
}

@article{mazin2021,
  title={Prediction of unconventional magnetism in doped {FeSb}$_2$},
  author={Mazin, I. I. and Koepernik, K. and Johannes, M. D. and Gonz{\'a}lez-Hern{\'a}ndez, R. and {\v{S}}mejkal, Libor},
  journal={Proceedings of the National Academy of Sciences of the United States of America},
  volume={118},
  pages={e2108924118},
  year={2021},
  publisher={National Academy of Sciences}
}

@article{yuan2021prm,
  title={Prediction of low-{Z} collinear and noncollinear antiferromagnetic compounds having momentum-dependent spin splitting even without spin-orbit coupling},
  author={Yuan, L.-D. and Wang, Z. and Luo, J.-W. and Zunger, Alex},
  journal={Physical Review Materials},
  volume={5},
  pages={014409},
  year={2021},
  publisher={American Physical Society}
}

@article{smejkal2022symmetry,
  title={Beyond conventional ferromagnetism and antiferromagnetism: {A} phase with nonrelativistic spin and crystal rotation symmetry},
  author={{\v{S}}mejkal, Libor and Sinova, Jairo and Jungwirth, Tomas},
  journal={Physical Review X},
  volume={12},
  pages={031042},
  year={2022},
  publisher={American Physical Society}
}

@article{bhowal2024,
  title={Ferroically ordered magnetic octupoles in d-wave altermagnets},
  author={Bhowal, Sayantika and Spaldin, Nicola A.},
  journal={Physical Review X},
  volume={14},
  pages={011019},
  year={2024},
  publisher={American Physical Society}
}

@article{mcclarty2024,
  title={Landau theory of altermagnetism},
  author={McClarty, P. A. and Rau, J. G.},
  journal={Physical Review Letters},
  volume={132},
  pages={176702},
  year={2024},
  publisher={American Physical Society}
}

@article{han2025octupoleHall,
  title={Harnessing magnetic octupole {H}all effect to induce torque in altermagnets},
  author={Han, Seungyun and Jo, Daegeun and Baek, Insu and Cheon, Suik and Oppeneer, Peter M. and Lee, Hyun-Woo},
  journal={Physical Review Letters},
  volume={135},
  number={7},
  pages={076705},
  year={2025},
  publisher={American Physical Society},
  doi={10.1103/lxkx-ypbg}
}

@article{tomiyoshi1982,
  title={Polarized Neutron Diffraction Study of the Spin Structure of {Mn}$_3${Sn}},
  author={Tomiyoshi, Shigeji},
  journal={Journal of the Physical Society of Japan},
  volume={51},
  number={3},
  pages={803--810},
  year={1982},
  publisher={The Physical Society of Japan},
  doi={10.1143/JPSJ.51.803}
}

@article{brown1990,
  title={Determination of the magnetic structure of {Mn}$_3${Sn} using generalized neutron polarization analysis},
  author={Brown, P. J. and Nunez, V. and Tasset, F. and Forsyth, J. B. and Radhakrishna, P.},
  journal={Journal of Physics: Condensed Matter},
  volume={2},
  number={47},
  pages={9409--9422},
  year={1990},
  publisher={IOP Publishing},
  doi={10.1088/0953-8984/2/47/015}
}

@article{nomoto2020,
  title={Cluster multipole dynamics in noncollinear antiferromagnets},
  author={Nomoto, Takuya and Arita, Ryotaro},
  journal={Physical Review Research},
  volume={2},
  pages={012045},
  year={2020},
  publisher={American Physical Society},
}

@MISC{fleurWeb,
  author = {},
  howpublished = {\url{https://www.flapw.de/}}
}

@article{yoon2023handedness,
  title={Handedness anomaly in a non-collinear antiferromagnet under spin–orbit torque},
  author={Yoon, Ju Young and Zhang, Pengxiang and Chou, Chung-Tao and Takeuchi, Yutaro and Uchimura, Tomohiro and Hou, Justin T. and Han, Jiahao and Kanai, Shun and Ohno, Hideo and Fukami, Shunsuke and Liu, Luqiao},
  journal={Nature Materials},
  volume={22},
  pages={1106--1113},
  year={2023},
  publisher={Nature Publishing Group},
}

@article{jackeli2009,
  title={Magnetically hidden order of {K}ramers doublets in d$_1$ systems: {Sr}$_2${VO}$_4$},
  author={Jackeli, G. and Khaliullin, G.},
  journal={Physical Review Letters},
  volume={103},
  pages={067205},
  year={2009},
  publisher={American Physical Society},
}

@article{iwazaki2023,
  title={Material-based analysis of spin-orbital {M}ott insulators},
  author={Iwazaki, Ryuta and Shinaoka, Hiroshi and Hoshino, Shintaro},
  journal={Physical Review B},
  volume={108},
  pages={L241108},
  year={2023},
  publisher={American Physical Society},
}

@article{perdew1996,
  title={Generalized gradient approximation made simple},
  author={Perdew, John P. and Burke, Kieron and Ernzerhof, Matthias},
  journal={Physical Review Letters},
  volume={77},
  pages={3865--3868},
  year={1996},
  publisher={American Physical Society},
}

@article{monkhorst1976,
  title={Special points for {B}rillouin-zone integrations},
  author={Monkhorst, Hendrik J. and Pack, James D.},
  journal={Physical Review B},
  volume={13},
  pages={5188--5192},
  year={1976},
  publisher={American Physical Society},
}

@inproceedings{zimmer1972,
  title={Investigation of the magnetic phase transformation in {Mn}$_3${Sn}},
  author={Zimmer, G. J. and Kr{\'e}n, E.},
  booktitle={AIP Conference Proceedings},
  volume={5},
  pages={513--516},
  year={1972},
  publisher={American Institute of Physics}
}

@article{mostofi2014,
  title={An updated version of wannier90: A tool for obtaining maximally-localised {W}annier functions},
  author={Mostofi, Arash A. and Yates, Jonathan R. and Pizzi, Giovanni and Lee, Young-Su and Souza, Ivo and Vanderbilt, David and Marzari, Nicola},
  journal={Computer Physics Communications},
  volume={185},
  number={8},
  pages={2309--2310},
  year={2014},
  publisher={Elsevier}
}

@article{pizzi2020wannier90,
  title={{W}annier90 as a community code: new features and applications},
  author={Pizzi, Giovanni and Vitale, Valerio and Arita, Ryotaro and Bl{\"u}gel, Stefan and Freimuth, Frank and G{\'e}ranton, Guillaume and Gibertini, Marco and Gresch, Dominik and Johnson, Charles and Koretsune, Takashi and Iba{\~n}ez-Azpiroz, Julen and Lee, Hyungjun and Lihm, Jae-Mo and Marchand, Daniel and Marrazzo, Antimo and Mokrousov, Yuriy and Mustafa, Jamal I. and Nohara, Yoshiro and Nomura, Yusuke and Paulatto, Lorenzo and Ponc{\'e}, Samuel and Ponweiser, Thomas and Qiao, Junfeng and Th{\"o}le, Florian and Tsirkin, Stepan S. and Wierzbowska, Ma{\l}gorzata and Marzari, Nicola and Vanderbilt, David and Souza, Ivo and Mostofi, Arash A. and Yates, Jonathan R.},
  journal={Journal of Physics: Condensed Matter},
  volume={32},
  pages={165902},
  year={2020},
  publisher={IOP Publishing},
}

@article{hayami2024unified,
  title={Unified description of electronic orderings and cross correlations by complete multipole representation},
  author={Hayami, Satoru and Kusunose, Hiroaki},
  journal={Journal of the Physical Society of Japan},
  volume={93},
  pages={072001},
  year={2024},
  publisher={Physical Society of Japan}
}

@article{shiina1997,
  title={Magnetic-field effects on quadrupolar ordering in a $\Gamma_8$-quartet system {CeB}$_6$},
  author={Shiina, Ryousuke and Shiba, Hiroyuki and Thalmeier, Peter},
  journal={Journal of the Physical Society of Japan},
  volume={66},
  number={6},
  pages={1741--1755},
  year={1997},
  publisher={Physical Society of Japan}
}

@article{kuramoto2009,
  title={Multipole orders and fluctuations in strongly correlated electron systems},
  author={Kuramoto, Yoshio and Kusunose, Hiroaki and Kiss, Annam\'aria},
  journal={Journal of the Physical Society of Japan},
  volume={78},
  pages={072001},
  year={2009},
  publisher={Physical Society of Japan}
}

@article{kusunose2008,
  title={Description of multipole in $f$-electron systems},
  author={Kusunose, Hiroaki},
  journal={Journal of the Physical Society of Japan},
  volume={77},
  pages={064710},
  year={2008},
  publisher={Physical Society of Japan}
}

@article{santini2009,
  title={Multipolar interactions in $f$-electron systems: {T}he paradigm of actinide dioxides},
  author={Santini, Paolo and Carretta, Stefano and Amoretti, Giuseppe and Caciuffo, Roberto and Magnani, Nicola and Lander, Gerard H.},
  journal={Reviews of Modern Physics},
  volume={81},
  pages={807--863},
  year={2009},
  publisher={American Physical Society}
}

@article{baek2025magoct,
  title={Magnetic octupole {H}all effect in heavy transition metals},
  author={Baek, Insu and Han, Seungyun and Lee, Hyun-Woo},
  journal={Physical Review B},
  volume={112},
  pages={064421},
  year={2025},
  publisher={American Physical Society},
  doi={10.1103/5hqs-v6y4},
  url={https://doi.org/10.1103/5hqs-v6y4}
}

@article{han2026deterministic,
  title={Deterministic N{\'e}el Vector Switching of Altermagnets Via Magnetic Octupole Torque},
  author={Han, S. and Baek, I. and Kim, K.-W. and Lee, H.-W. and Cheon, S.},
  journal={Small},
  volume={22},
  number={23},
  pages={e11790},
  year={2026},
  doi={10.1002/smll.202511790}
}

\end{document}